\documentclass[letterpaper,12pt]{article}
\usepackage[top=1in,bottom=1in,left=1.5in,right=1in]{geometry}
\usepackage{amssymb,amsmath} 
\usepackage{graphicx} 
\usepackage{color} 
\usepackage{appendix}
\usepackage{longtable}
\usepackage{afterpage}

\usepackage{titlesec}
\usepackage{authblk}
\usepackage{setspace}
\titleformat{\section}[hang]{\LARGE\bfseries}{}{1em}{}
\titleformat{\subsection}[hang]{\bfseries}{}{1.2em}{}
\titleformat{\subsubsection}[hang]{\itshape}{}{1.5em}{}
\usepackage[square, comma, sort&compress, numbers]{natbib}
\begin{document}
\title{A MATLAB Program for Quantitative Simulation of Self-assembly of 
Polymer Blend Films with Nano-scaled Features}

\author{Yingrui Shang\footnote{ys2503@columbia.edu}}
\author{David Kazmer}

\affil{Center of High-rate Nano-manufacturing University of Massachusetts 
Lowell 2008} 

\maketitle

\clearpage

\abstract
A MATLAB program has been developed for simulation of polymer blend 
self-assembly with nano-scaled features. The Cahn-Hilliard equation is 
implemented to calculate the free energy profile of the polymer blends. The 
Flory-Huggins type of energy is used to estimate the local free energy. The 
program is capable of quantitatively simulate the phase separation of polymer
blends. The effects such as the substrate functionalization, solvent 
evaporation, and polymer materials properties are included in the program. The 
program can estimate the model parameters from the real experimental processing
parameters and the material properties. The simulation results can be evaluated 
quantitatively and compared with the experimental results with analysis 
tools included in the program.

\section{Introduction}
Nano-manufacturing by polymer self-assembly is attracting interests in recent 
decades due to its wide applications~\cite{FINK:1998}. The numerical simulation
of this process can be used to research the mechanisms of phase separation of 
polymer blends and predict the unobservable process states and unmeasurable 
material properties. The mathematical principles and numerical simulation of 
self-assembly via phase separation has been extensively 
studied~\cite{SCOTT:1949,HSU:1973,CHEN:1994,HUANG:1995,ALTENA:1982,ZHOU:2006,TONG:2002,HE:1997,MUTHUKUMAR:1997,KARIM:1998}. But few specific software
toolkit have been developed to efficiently investigate this phenomenon. \par

A computer program is developed in MATLAB for the numerical simulation of the 
polymer blends phase separation. 
With this software, the mechanisms of the phase separation are investigated. 
Also the mobility, gradient energy coefficient energy, and the surface energy 
in the experiment are estimated with the numerical model. The software can 
evaluate the physical parameters in the numerical model by implementing the 
real experimental parameters and materials properties. The numerical simulation
results can be analyzed with the software and the results from the simulation
software can be validated with the experimental results. \par

\begin{figure}[!ht]
	\centering
	\includegraphics[width=\textwidth]{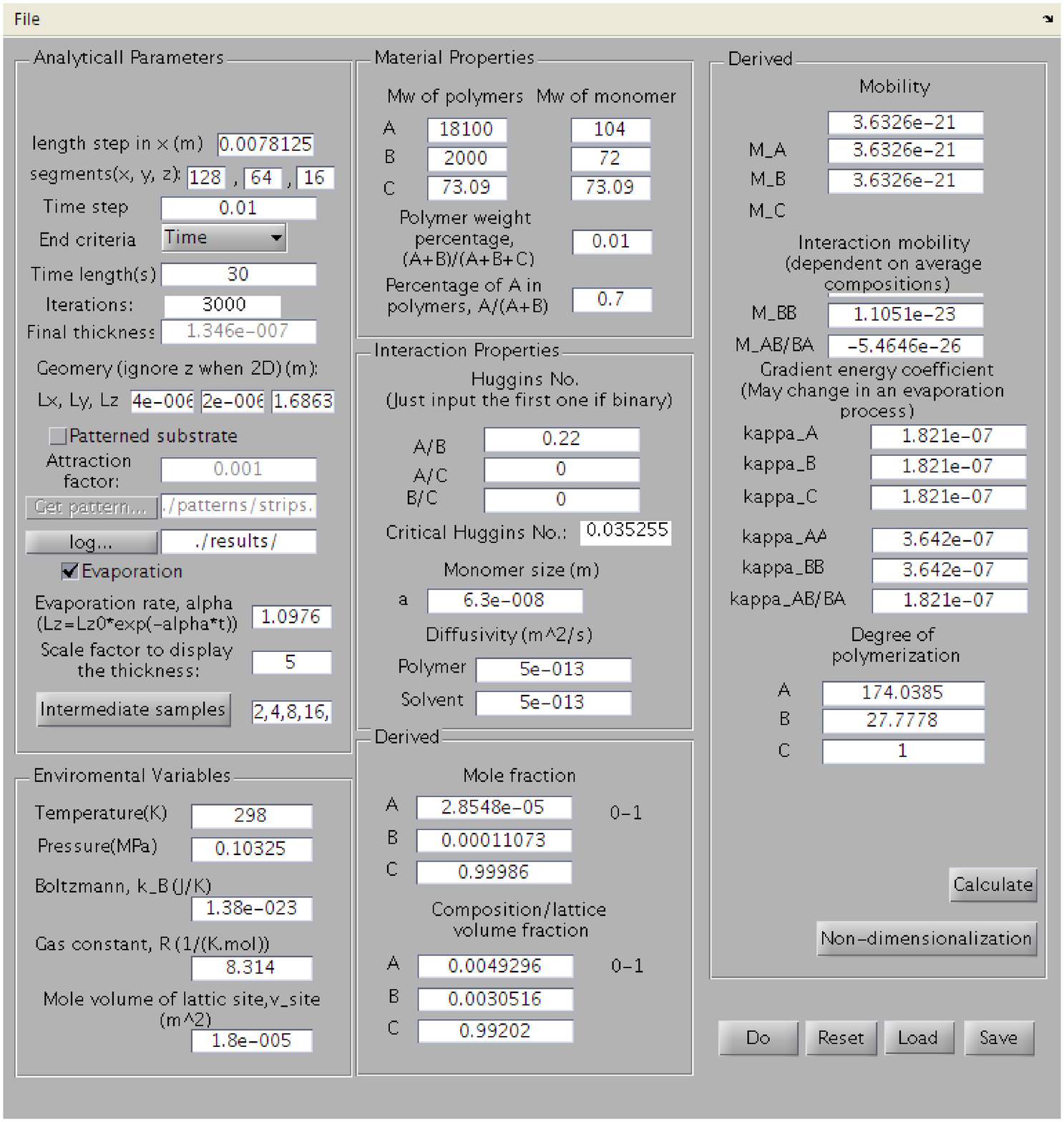}
	\caption{Screenshot of the simulation program graphical user interface.
	\label{fg_gui_screenshot}}
\clearpage
\end{figure} 

\section{Fundamentals}
The numerical model for phase separation of polymer blends are established and 
validated with experimental results work~\cite{SHANG:2010}. The free energy profile during the phase separation in a inhomogeneous mixture 
is described by Cahn-Hilliard 
Equation~\cite{CAHN:1958, CAHN:1959, CAHN:1961, CAHN:1965}, as shown below,

\begin{equation}
	F(C_1,C_2,C_3)=\int_{V} \left\{ f(C_1,C_2,C_3)+\displaystyle\sum_{i=1,2,3} [\kappa_i (\nabla C_i)^2] \right\} dV \label{cahn_hilliard_intro}
\end{equation}

where $f$ is the local free energy density of homogeneous material, $\phi _i$ 
is the lattice volume fraction of component $i$, and $\kappa_i$ is the gradient
energy coefficient for the component $i$. The total free energy of the system 
is composed by two items as shown in Equation~\ref{cahn_hilliard_intro}. The 
first item is the local free energy and the second is the composition gradient 
contribution to the free energy. \par

In our study, the local free energy is in the form of Flory-Huggins equation,
which is well know and studied for polymer blends~\cite{HUANG:1999}
The ternary Flory-Huggins Equation is shown as follows, 

\begin{equation}
	\begin{split}
		f(C_1,C_2,C_3)
			 &=  \frac{RT}{v_{site}}\bigg( \frac{C_1}{m_1}\ln{C_1}+\frac{C_2}{m_2}\ln{C_2} + C_3\ln{C_3} \\
			& \chi_{12}C_1C_2+\chi_{13}C_1C_3+\chi_{23}C_2C_3\bigg) 
\label{eq_flory_huggins_intro}
	\end{split}
\end{equation}

where $R$ is the ideal gas constant, $T$ is the absolute temperature, 
$v_{site}$ is the lattice site volume in the Flory-Huggins model, $m_i$ is the 
degree of polymerization of component $i$, and $C_i$ is the composition for the
component $i$. \par

There are some parameters in the numerical model which can not be measured
directly, such as the gradient energy coefficient and the mobility. These 
parameters have to be estimated from the experimental 
parameters.The gradient energy coefficient, $\kappa$, determines the influence 
of the composition gradient to the total free energy of the domain. 
The value of $\kappa$ is difficult to measure experimentally. Though efforts 
have been made by Saxena and Caneba~\cite{SAXENA:2002} to estimate the 
gradient energy coefficient in a ternary polymer system from experimental 
methods, few experimental results are published for our conditions. Initially, 
the value of $\kappa$ can be estimated by the interaction distance between 
molecules~\cite{WISE_THESIS:2003},

\begin{equation}
	\kappa=\frac{RTa^2}{3v_{site}}\label{eq_gradient_energy_coefficient}
\end{equation} 

where $a$ is the monomer size. A modified equation to calculate $\kappa$ 
considering the effects of the composition is reported by Gennes, 
et al.~\cite{GENNES:1980}.

\begin{equation}
	\kappa_i=\frac{RTa^2}{36v_{site}C_i}
\end{equation} 

where the subscript, $i$, represents component $i$.  \par

The mobility is estimated from the diffusivity of the components. The mobility 
of the polymer blends with long chains can be estimated by the equation as 
follows~\cite{GENNES:1980}, 

\begin{equation}
M_i=\frac{C_i}{m_i}\frac{D_mN_ev_{site}}{RT}
\end{equation} 

where $m_i$ is the degree of polymerization as stated before, $D_m$ is the 
diffusivity of the monomer, and $N_e$ is the effective number of monomers per 
entanglement length. Because of the scarce experimental data for $N_e$, a more 
generalized form is employed for our study,  

\begin{equation}
M=\frac{Dv_{site}}{RT}\label{eq_mobility}
\end{equation} 

The time evolution of the composition of component $i$ can be represented 
as~\cite{HUANG:1995,BATTACHARYYA:2003,GENNES:1980,SHANG:2009},\par

\begin{equation}
	\begin{split}
		\frac{\partial C_i}{\partial t}
		&= M_{ii}\left[ \frac{\partial f}{\partial C_i}-\frac{\partial f}{\partial C_3}-2\kappa_{ii}\nabla^2C_i-2\kappa_{ij}\nabla^2C_j\right] \\
		& +M_{ij}\left[ \frac{\partial f}{\partial C_j}-\frac{\partial f}{\partial C_3}-2\kappa_{ji}\nabla^2C_i-2\kappa_{jj}\nabla^2C_j \right]
	\end{split}\label{eq6_paper2}
\end{equation}

where the subscripts $i$ and $j$ represent components 1 and 2, and\par

\begin{equation}
	\begin{aligned}
		M_{ii}=&(1-\overline{C}_i)^2M_i+\overline{C}_i^2\displaystyle\sum_{j\neq i}M_j\qquad i=1,2;j=1,2,3\\
		M_{ij}=&-\displaystyle\sum_{i\neq j}\left[(1-\overline{C}_i)\overline{C}_j\right]M_i+\overline{C}_i\overline{C}_jM_3\qquad i=1,2;j=1,2
	\end{aligned}
\end{equation}

where $\overline{C}_i$ is the average composition of component $i$. To simplify
the solution of Equation \ref{eq6_paper2}, $\kappa_{ii}=\kappa_i+\kappa_3$, and
$\kappa_{12}=\kappa_{21}=\kappa_3$, where $\kappa_i$ is the gradient energy 
coefficient in Equation~\ref{eq_gradient_energy_coefficient}. \par

For detailed discussion and practical scientific cases with this software can 
be found in our previous 
works~\cite{SHANG:2008,SHANG:2009,SHANG:2009THESIS}.\par

\section{The MATLAB Program for Simulation of Polymer Phase Separation}

\subsection{Design Principles}
The program is developed in MATLAB m-code. A graphical user interface (GUI) is 
implemented in the program created with MATLAB GUI editor. MATLAB is widely 
used in scientific computation and has many toolkits and commonly used 
mathematical functionalities. But implementing the software in MATLAB the
efficiency of development is greatly improved. Also, by developing the program
in MATLAB, the program is cross platform. \par
  
The software is designed for daily usage of simulation and experiment 
scientists. The program is light weighted and programmed with high computation
efficiency so that it can produce significant science results in a common PC.
It also extensible to a parallel version or implement code to use the high 
computation performance of GPU. The GUI is implemented so that the users can 
conveniently input the experiment parameters. The results as well as the user
settings can be saved and revisited by the program. Also, for better assistance
to a real productive environment, the simulation model is carefully designed, 
so that the users provide the real processing and material parameters and the 
program will produce quantitative results comparable to experimental results.
Analytical tools are also provided with the program for post-processing of the
results. \par

\subsection{Numerical Methods}
To solve the partial differential equation, the discrete cosine transform 
spectral method is employed. The discrete cosine transform (DCT) is applied 
to the right hand side and left hand side of Equation~\ref{eq6_paper2}. The 
partial differential equation in the ordinary space then transformed into an 
ordinary differential equation in frequency space. When the ODE in frequency 
space then is solved, the results are transformed back to the ordinary 
space. \par

Comparing to conventional finite element method, the spectral method is more 
efficient and accurate. This method enabled the program to solve the equation 
in a reasonable computation time to investigate the changes of the phase 
separation during an real time span long enough to observe the phase evolution. 
The spectral method is only
applied to the spatial coordinates since the time length of the evolution is
not predictable. Actually the real time for phase evolution is usually one of
the major concerns as the result of the simulation. \par 

The DCT takes a considerable portion of the computation time. Especially in a 
3-dimensional numerical model, the 3-dimensional DCT function with conventional
approach has a complexity of $O(n^3)$, which can not be practical for real 
application on a PC. To overcome this computational difficulty, the code can 
either be translated to C code embedded in MATLAB m scripts, or a different 
mathematical approach can be implemented as well. In this program, the DCT is 
calculated from the fast Fourier transform (FFT) which is optimized in 
MATLAB. \par

\subsection{Quantitative Simulation with Real Experimental Parameters}
Many of previous numerical simulations in the self-assembly with polymer blends
phase separation are qualitative other than quantitative. The results can only 
be used to provide non-quantitative suggestions to the experiments. While this
program implemented a numerical model which quantitatively simulates the 
experimental results with the real processing and material parameters. Most of 
inputs in to this program can be directly measured and read from the instrument
or material labels. For some of the physical parameters such as $\kappa$ and
the mobility, the program can provide a start value from the calculation with
the theoretical model. The user may need to validate the value by comparing 
the simulation results to the experimental results. Eventually, a more accurate 
estimation can be find with optimization methods by setting the difference 
between the simulation and experiment results as the cost function. \par 

Besides the parameters in Cahn-Hilliard equation, other effects such as the 
evaporation, substrate functionalization, and the degree of polymerization are
also implemented with the real conditions. The final results are saved and 
summarized. The characteristic length of result pattern from simulation and its
compatibility with the substrate functionalization are calculated. These 
numbers can be used to compare with the experimental results. \par

\subsection{Data Visualization and Results Analysis}    
When running the program, the message from the software will be output to the 
working console of MATLAB. The messages will show the current state and real
time results of the simulation. Also, when the simulation is started, the phase 
pattern will be plotted in a real time plot window. Users can set the frequency
of real time plot and the scale factor on the domain of the contour plot in
the GUI. The results of the simulation will be saved to a folder designated by
the user. The real time plot will be saved to the result folder. The 
quantitative results will be saved as several comma separated values (CSV)
text files. The result folder can be loaded into the analysis toolkit of the 
program and the user can view the assessment values such as the characteristic
length, the compatibility parameters, and the composition profile wave in depth
direction with convenient plotting tools. Usually these results such as the 
composition profile in each direction in the domain are difficult to observe
in experiment results. \par

\begin{figure}[!ht]
	\centering
	\includegraphics[width=\textwidth]{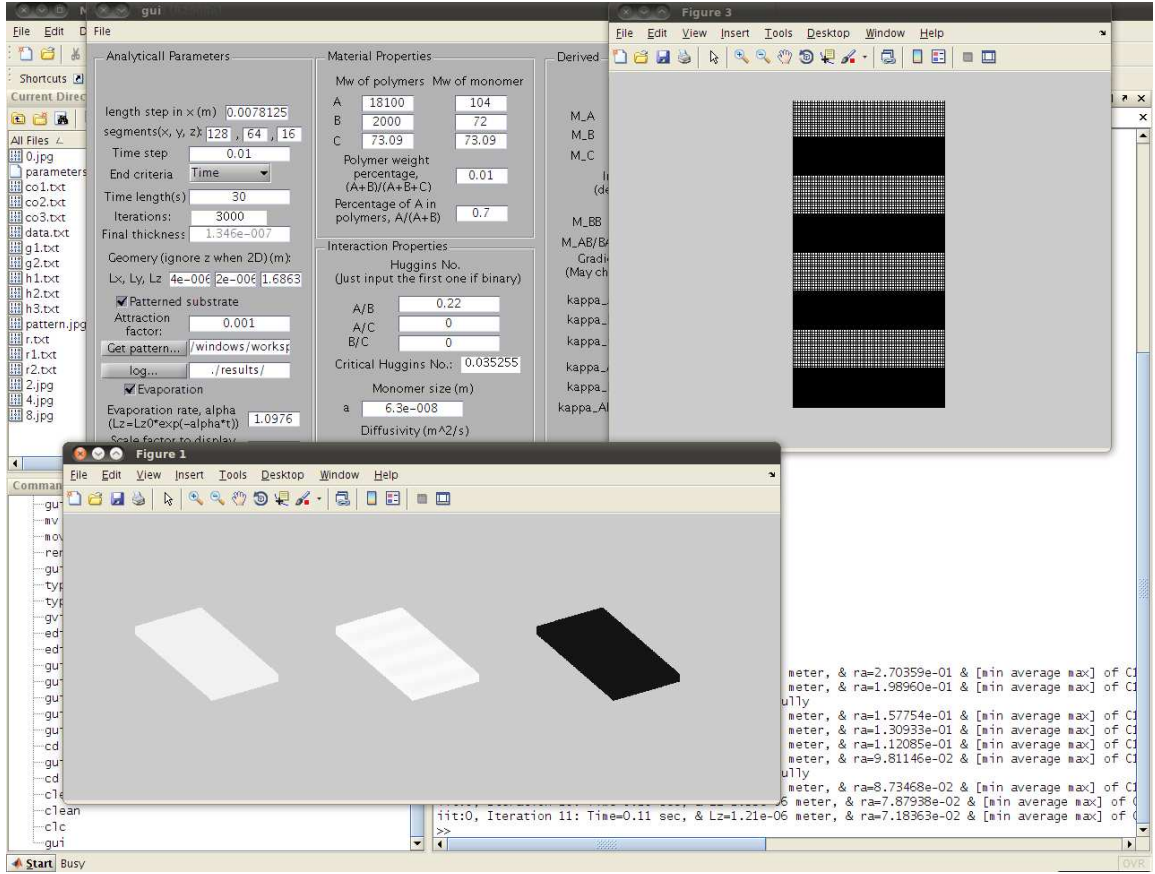}
	\caption{The simulation is running with the real time plot of the 
	current ternary phase morphology.
	\label{fg_gui_running}}
\clearpage
\end{figure}

\section{Examples}
To demonstrate the capability of this program, example simulation cases are 
shown in this paper. The results of numerical simulation have been validated
with the experimental results in our previous work~\cite{SHANG:2010}. To 
compare the simulated results with a real experimental system, we directed 
the morphologies of polystyrene (PS) / polyacrylic acid (PAA) blends using 
chemically heterogeneous patterns.  More specifically, alkanethiols with 
different chemical functionalities were patterned by electron beam 
lithography, which were then used to direct the assembly of PS/PAA blends 
during the spin coating from their mutual solvent~\cite{MING:2009}. The 
experimental conditions are implemented into the numerical simulation. The
effects such as the substrate functionalization and the solvent evaporation
are involved in the numerical modeling. 
The parameters difficult to measure are acquired with the optimization methods
~\cite{SHANG:2009}.   
\par

\begin{figure}[!ht]
    \centering
    \includegraphics[width=\textwidth]{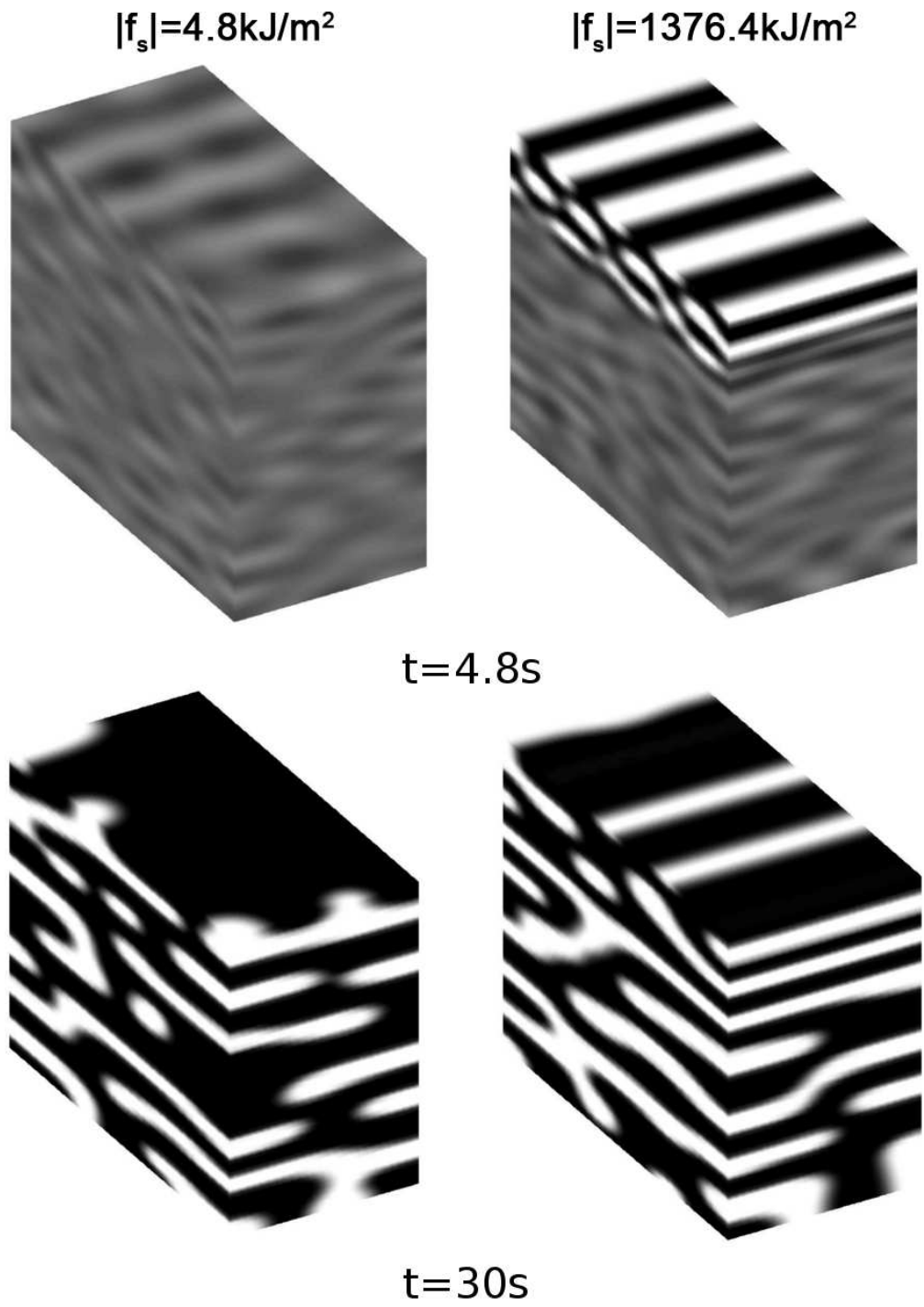}
    \caption{The phase separation in a thick film. \label{fg_thick_film}}
\clearpage
\end{figure}

Sophisticated techniques are required to investigate the composition profile in
the depth of the polymer film~\cite{GEOGHEGAN:2003}. While the numerical 
simulation results can provide the composition profile in each position of the 
file, the composition profile change in depth direction can be easily accessed.
To investigate the composition wave allow the direction perpendicular to the 
film surface, a thick film is implemented to the numerical simulation. This kind
of film is not only difficult to fabricate and characterize in experiments, however
in the numerical modeling, the user only needs to change the mesh grid domain size. 
The depth profiles with different substrate functionalization are shown in  
Figure~\ref{fg_thick_film}, where $|f_s|$ denotes the surface energy term from the 
substrate functionalization. This term will be added to the total free energy 
on the interface of the polymer film and the substrate. The initial thickness
of the film is 1 mm and decreases to 8 $\mu m$ due to the evaporation of the 
solvent. The thickness results are scaled by 0.5 to fit in the figures. It can 
be seen that a higher surface interaction force can result in a faster substrate
directed phase separation in the film. A stronger substrate interface attraction
force can direct the phase separation morphology near the substrate surface. While
with a lower surface energy, the phase separation dynamics in the bulk of the 
film overcomes the substrate attraction force. It can be seen that at 30 seconds,
the substrate functionalization has little effects on the morphology on the
substrate surface. Also, the checker board structure can be seen near the 
substrate surface with a higher surface energy~\cite{KARIM:1998}. \par

\begin{figure}[!ht]
	\centering
	\includegraphics[width=\textwidth]{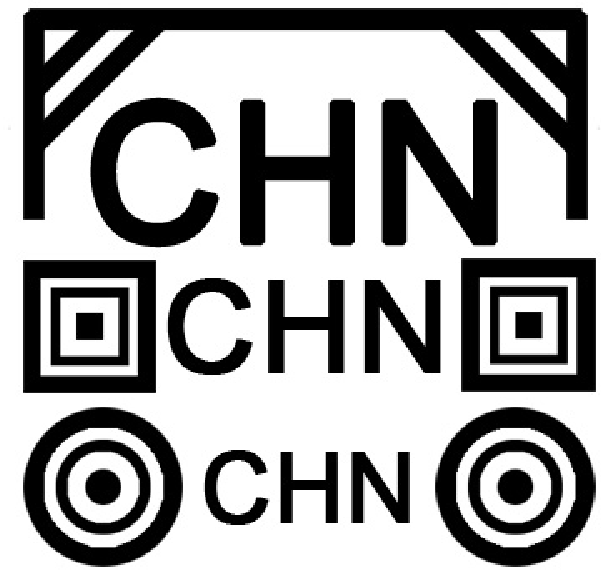}
	\caption{The substrate pattern with complicated features. 
	\label{fg_chn_pattern}}
\clearpage
\end{figure}

\begin{figure}[!ht]
	\centering
	\includegraphics[width=\textwidth]{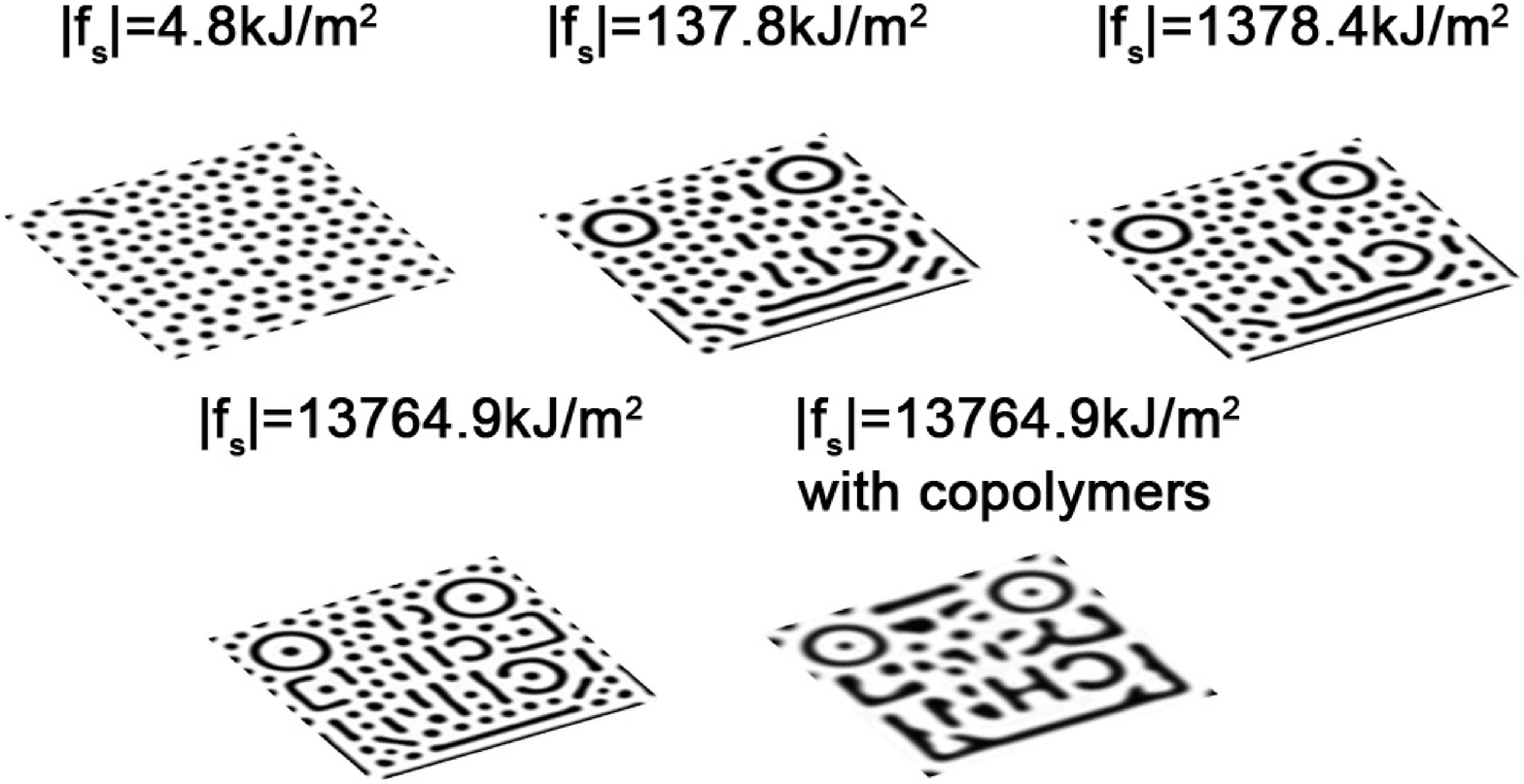}
	\caption{The effects of complicated substrate patterns. 
	\label{fg_complicated_patterns}}
\clearpage
\end{figure}

To investigate the effects of a more complicated pattern, a larger domain is 
simulated. The pattern on the substrate applied on the substrate surface is 
shown in Figure~\ref{fg_chn_pattern}. The substrate pattern is designed to 
investigate the effects of various shapes and contains components such as 
squares, circles, and dead end lines in different sizes. The initial surface 
dimensions of the model are changed to 12$\mu m\times$12$\mu m$. The initial 
thickness of the film is 1mm and shrinks during the solvent evaporation. The 
elements in the modelling is 384$\times$384$\times$16. The average composition 
ratio of PS/PAA is changed to 38/62 to match the pattern. The result patterns 
from the simulation can be seen in Figure~\ref{fg_complicated_patterns}. \par

\begin{figure}[!ht]
	\centering
	\includegraphics[width=\textwidth]{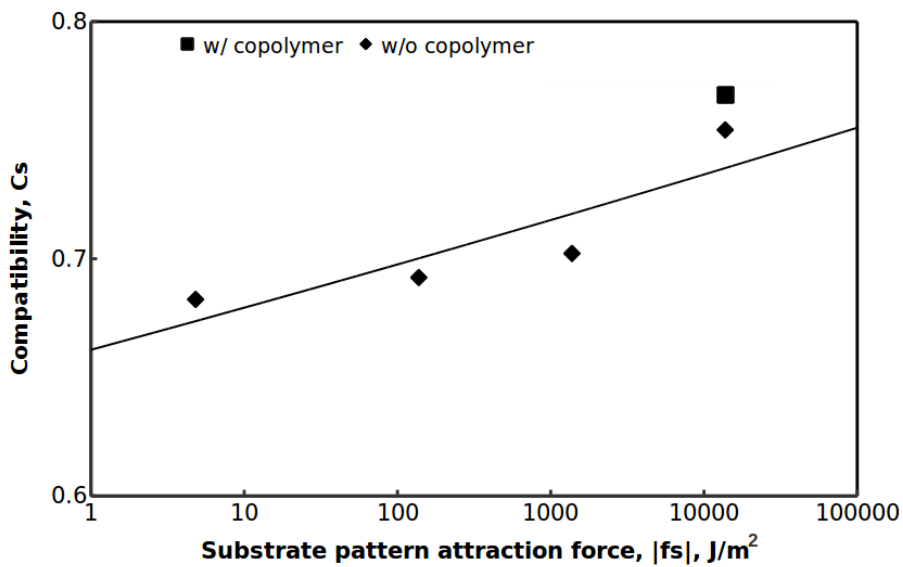}
	\caption{The effects of complicated substrate patterns. \label{fg_cs_complicated_patterns}}
\clearpage
\end{figure}

It can be seen that in a larger domain with complicated substrate patterns, the
attraction factor has to be increased to obtain a better replication. In 
general, the increase of the attraction factor will increase the refinement of 
the pattern according to the substrate pattern. But since the substrate pattern
has geometrical features in different sizes, the attraction factor has to be 
strong enough to force the intrinsic phase separation with unified 
characteristic length to match the substrate pattern in different sizes. This 
would be the main challenge to the replication of complicated patterns. It has 
been reported by Ming et. al.~\cite{MING:2009} that the addition of the 
copolymer can improve the refinement of the final patterns in experiments. The 
reason is that the PAA-b-PS block copolymer will concentrate in the interface 
of the PS and PAA domains in the phase separation, therefore decreasing the 
mixing free energy. Fundamentally, the addition of the block copolymer 
increased the miscibility of the two polymers. To simulate these phenomena, the
Flory-Huggins interaction parameter is decreased from 0.22 to 0.1 to increase 
the miscibility of PS/PAA in the modelling. The result pattern is also shown in
Figure~\ref{fg_complicated_patterns}, in comparison to the cases without the 
addition of block copolymers. It can be seen that the refinement of the phase 
separated pattern is improved by the addition of the block copolymer. The $C_s$
values of the phase separation with complicated patten are measured and plotted
in Figure~\ref{fg_cs_complicated_patterns}. \par

A assessment parameter, $C_s$, the compatibility parameter is introduced to 
evaluate the replication of the morphology to the substrate pattern, where 
a higher $C_s$ value denotes a better 
replication of the polymer film morphology according to the substrate pattern
It can be seen in Figure~\ref{fg_cs_complicated_patterns} that the $C_s$ value 
for the system with block copolymer is 7.69E-01, which is higher than the 
system without the block copolymer when attraction forces are the same. The 
decrease of the Flory-Huggins interaction parameter increases the miscibility 
of the polymers, which will decrease the miscibility gap of the polymers, as 
can be seen in Equation~\ref{eq_flory_huggins_intro}. The two phase at 
equilibrium will be less concentrated in different types of polymer. This is an
issue may need to be concerned when the interaction parameter of the two 
polymers is changed. \par

\section{Conclusion}

A computer program for simulation of polymer self-assembly with phase separation
is introduced. The program is developed in MATLAB m code and designed to
assist the scientists in real working environments. The program is able to 
simulate the experiment results quantitatively with real experimental
parameters. The unmeasurable physical parameters such as the gradient energy
coefficient and the mobility can be estimated with the program. The program 
provides a graphical user interface and analytical toolkits. This program 
can help the scientists in research in polymer phase separation mechanisms and 
dynamics with high efficiency, convenience of usage, quantitative results 
analysis, and validated reliability. 

\section{Acknowledgement}
The authors would thank the efforts of Liang Fang and Ming Wei for providing
help in the experimental procedurals. The authors also appreciate the
valuable suggestions and comments from other users and testers of this program.
This project is a part of the research in Center of High-rate Nanomanufacturing,
sponsored by National Science Foundation (grant number NSF-0425826).

\bibliographystyle{unsrt} 
\bibliography{abb-journal-names,polymer-blend-self-assembly}    

\begin{thebibliography}{10}

\bibitem{FINK:1998}
Yoel Fink, Joshua~N. Winn, Shanhui Fan, Chiping Chen, Jurgen Michel, John~D.
  Joanopoulos, and Edwin~L. Thomas.
\newblock A dielectric omnidirectional reflector.
\newblock {\em Science}, 282(5394):1679--1682, 1998.

\bibitem{SCOTT:1949}
Robert~L. Scott.
\newblock The thermodynamics of high polymer solutions. v. phase equilibria in
  the ternary system: Polymer 1-polymer 2-solvent.
\newblock {\em J. Chem. Phys.}, 17(3):279--284, 1949.

\bibitem{HSU:1973}
C.~C. Hsu and J.~M. Prausnitz.
\newblock Thermodynamics of polymer compatibility in ternary systems.
\newblock {\em Macromolecules}, 7(3):320--324, 1973.

\bibitem{CHEN:1994}
Long-qing Chen.
\newblock Computer simulation of spinodal decomposition in ternary systems.
\newblock {\em Acta Matall. Mater.}, 42(10):3503--3513, 1994.

\bibitem{HUANG:1995}
C.~Huang, M.~Olvera de~la Cruz, and B.~W. Swift.
\newblock Phase separation of ternary mixtures: Symmetric polymer blends.
\newblock {\em Macromolecules}, 28:7996--8005, 1995.

\bibitem{ALTENA:1982}
Frank~W. Altena and C.~A. Smolders.
\newblock Calculation of liquid-liquid phase separation in a ternary system of
  a polymer in a mixture of a solvent and a nonsolvent.
\newblock {\em Macromolecules}, 15:1491--1497, 1982.

\bibitem{ZHOU:2006}
Bo~Zhou and Adam Powell.
\newblock Phase field simulations of liquid-liquid demixing during immersion
  precipitation of polymer membranes in 2d and 3d.
\newblock {\em J. Membrane Sci.}, 268(2):150--164, 2006.

\bibitem{TONG:2002}
Chaohui Tong, Hongdong Zhang, and Yuliang Yang.
\newblock Phase separation dynamics and reaction kinetics of ternary mixture
  coupled with interfacial chemical reaction.
\newblock {\em J. Phys. Chem.~B}, 106:7869--7877, 2002.

\bibitem{HE:1997}
David~Qiwei He and E.~B. Nauman.
\newblock Spinodal decomposition with varying chain lengths and its application
  to designing polymer blends.
\newblock {\em J. Polym. Sci. Pol. Phys.}, 35(6):897--907, 1997.

\bibitem{MUTHUKUMAR:1997}
M.~Muthukumar, C.~K. Ober, and E.~L. Thomas.
\newblock Competing interaction and levels of ordering in self-organizing
  poymer materials.
\newblock {\em Science}, 277:1225--1232, 1997.

\bibitem{KARIM:1998}
A.~Karim, J.~F. Douglas, B.~P. Lee, S.~C. Glotzer, J.~A. Rogers, R.~J. Jackman,
  E.~J. Amis, and G.~M. Whitesides.
\newblock Phase separation of ultrathin polymer-blend films on patterned
  substrates.
\newblock {\em Phys. Rev.~E.}, 57(6):6273--6276, 1998.

\bibitem{SHANG:2010}
Yingrui Shang, Liang Fang, David Kazmer, Ming Wei, Carol Barry, and Joey Mead.
\newblock Verification of numerical simulation of the self-assembly of
  polymer-polymer-solvent ternary blends on a heterogeneously functionalized
  substrate.
\newblock {\em Macromolecules}, submitted, 2010.

\bibitem{CAHN:1958}
John~W. Cahn and John~E. Hilliard.
\newblock Free energy of a nonuniform system. i. interfacial free energy.
\newblock {\em J. Chem. Phys.}, 28(2):258--267, 1958.

\bibitem{CAHN:1959}
John~W. Cahn.
\newblock Free energy of a nonuniform system. ii. thermodynamic basis.
\newblock {\em J. Chem. Phys.}, 30(5):1121--1124, 1959.

\bibitem{CAHN:1961}
J.~W. Cahn.
\newblock On spinodal decomposition.
\newblock {\em Adv. Funct. Mater.}, 9:795--801, 1961.

\bibitem{CAHN:1965}
John~W. Cahn.
\newblock Phase separation by spinodal decomposition in isotropic systems.
\newblock {\em J. Chem. Phys.}, 42(1):93--99, 1965.

\bibitem{HUANG:1999}
C.~Huang, M.~Olvera De~La Cruz, and P.~W. Voorhees.
\newblock Interficial adsorption in ternary alloys.
\newblock {\em Acta Mater.}, 47(17):4449--4459, 1999.

\bibitem{SAXENA:2002}
R.~Saxena and G.~T. Caneba.
\newblock Study of spinodal decomposition in a ternary
  polymer-solvent-nonsolvent system.
\newblock {\em Poly. Eng. Sci.}, 2002.

\bibitem{WISE_THESIS:2003}
Steven~M. Wise.
\newblock {\em Diffuse Interface Model for Microstructural Evolution of
  Stressed, Binary Thin Films on Patterned Substrates}.
\newblock PhD thesis, University of Virginia, 2003.

\bibitem{GENNES:1980}
P.~G.~de Gennes.
\newblock Dynamics of fluctuations and spinodal decomposition in polymer
  blends.
\newblock {\em J. Chem. Phys.}, 72(9):4756--4763, 1980.

\bibitem{BATTACHARYYA:2003}
Saswata Battacharyya and T.~A. Abinandanan.
\newblock A study of phase separation in ternary alloys.
\newblock {\em Bull. Mater. Sci.}, 26:193--197, 2003.

\bibitem{SHANG:2009}
Yingrui Shang, David Kazmer, Ming Wei, Joey Mead, and Carol Barry.
\newblock Numerical simulation of the self-assembly of a
  polymer-polymer-solvent ternary system on a heterogeneously functionalized
  substrate.
\newblock {\em Poly. Eng. Sci.}, accepted, 2009.

\bibitem{SHANG:2008}
Yingrui Shang, David Kazmer, Ming Wei, Joey Mead, and Barry Carol.
\newblock Numerical simulation of phase separation of immiscible polymer blends
  on a heterogeneously functionalized substrate.
\newblock {\em J. Chem. Phys.}, 128(22), 2008.

\bibitem{SHANG:2009THESIS}
Shang Yingrui.
\newblock {\em Numerical Simulation for the Self-assembly of Polymer Blends
  with Nano-scaled Features}.
\newblock PhD thesis, University of Massachusetts at Lowell, 2009.

\bibitem{MING:2009}
Ming Wei, Liang Fang, Jun Lee, Sivasubramanian Somu, Xugang Xiong, Carol Barry,
  Ahmed Busnaina, and Joey Mead.
\newblock Directed assembly of polymer blends by self assembly of alkanethiols
  on electon beam lithography patterned templates.
\newblock {\em Adv. Mater.}, 21:794--798, 2009.

\bibitem{GEOGHEGAN:2003}
Mark Geoghegan and Georg Krausch.
\newblock Wetting at polymer surface and interfaces.
\newblock {\em Prog. Polym. Sci.}, 28:261--302, 2003.

\end{thebibliography}

\end{document}